# Theory for the Accuracy of Microcomb Photonic Microwave Transversal Signal Processors

*David J. Moss*

*Abstract*—**Photonic RF transversal signal processors, which are equivalent to reconfigurable electrical digital signal processors but implemented with photonic technologies, have been widely used for modern high-speed information processing. With the capability of generating large numbers of wavelength channels with compact micro-resonators, optical microcombs bring new opportunities for realizing photonic RF transversal signal processors that have greatly reduced size, power consumption, and complexity. Recently, a variety of signal processing functions have been demonstrated using microcomb-based photonic RF transversal signal processors. Here, we provide detailed analysis for quantifying the processing accuracy of microcomb-based photonic RF transversal signal processors. First, we investigate the theoretical limitations of the processing accuracy determined by tap number, signal bandwidth, and pulse waveform. Next, we discuss the practical error sources from different components of the signal processors. Finally, we analyze the contributions of the theoretical limitations and the experimental factors to the overall processing inaccuracy both theoretically and experimentally. These results provide a useful guide for designing microcomb-based photonic RF transversal signal processors to optimize their accuracy.**

## I. INTRODUCTION

After decades of development, the use of electronic devices for signal processing is rapidly approaching intrinsic limitations for processing bandwidth [1, 2]. In contrast, optical signal processing technologies can overcome this bandwidth bottleneck to provide processing speeds orders of magnitude faster [3, 4], critical for high-speed information processing applications. In the past two decades, a variety of signal processing functions have been realized based on different types of optical devices and systems, such as fiber gratings [5, 6], semiconductor optical amplifiers (SOAs) together with optical filters [7, 8], meta-surfaces in free-space optical systems [9-11], RF photonic systems with integrated micro-resonators or subwavelength gratings serving as optical processing modules [3, 12-14], and photonic RF transversal signal processing systems [15-18]. Amongst them, photonic

RF transversal signal processing systems show competitive advantages in achieving highly reconfigurable signal processing based on a single system [19, 20], which are attractive for meeting different processing requirements in practical applications.

To achieve a high processing accuracy, photonic RF transversal signal processors require many wavelength channels to serve as discrete taps to sample the input RF signal. Although conventional multi-wavelength sources such as discrete laser arrays [21-23] and fiber Bragg grating arrays [24-26] have been used to provide the discrete taps, they suffer from being limited in the tap numbers they can provide given that the system size, power consumption, and complexity increase dramatically with the tap number. In contrast, optical microcombs [27, 28], which are laser frequency combs (LFCs) generated by micro-resonators with high quality (Q) factors, show distinctive advantages by simultaneously providing large numbers of wavelength channels based on a single compact device. In addition, compared to LFCs generated by mode-locked fiber lasers [29, 30], optical microcombs have large comb spacings enabled by the small volume of the micro-resonators, which yield wide Nyquist bands between different wavelength channels that allow for large operation bandwidths for the transversal signal processors [19, 20]. Recently, a variety of signal processing functions have been demonstrated using microcomb-based photonic RF transversal signal processors, including not only basic processing functions such as differentiation [16, 31], integration [17], and Hilbert transform [15, 18], but also more complex functions such as phase encoding [32], arbitrary waveform generation [33, 34], and computing in optical neural networks [35-37].

David J. Moss is with the Optical Sciences Center, Swinburne University of Technology, Hawthorn, VIC 3122, Australia (e-mail: dmoss@swin.edu.au).

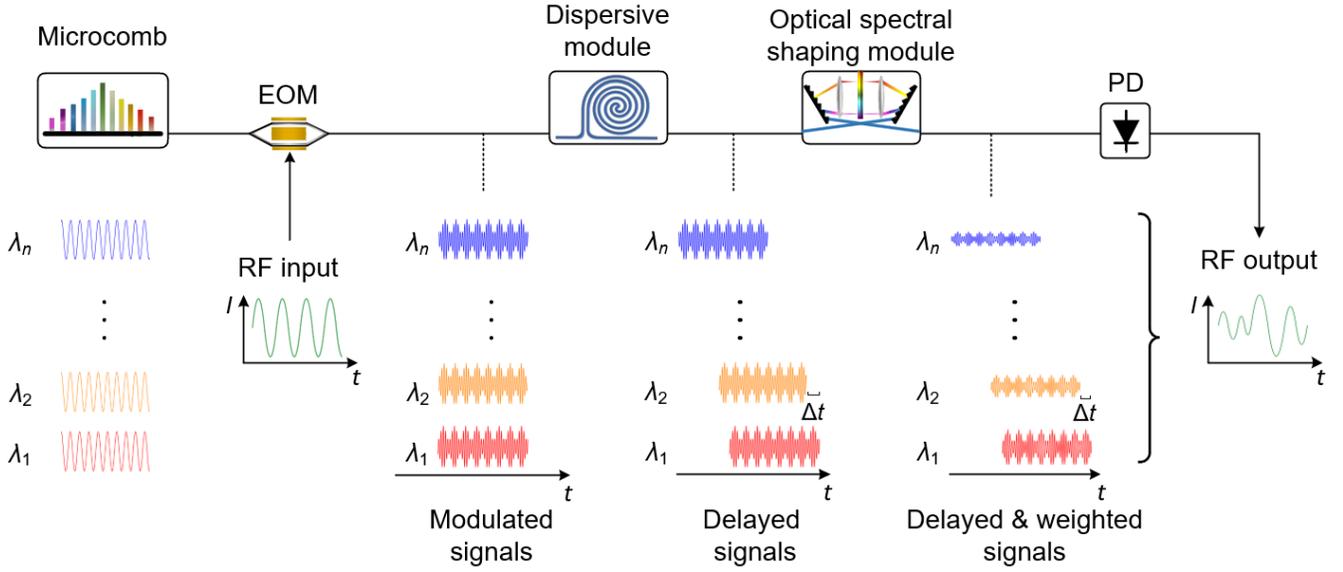

**Fig. 1.** Schematic diagram and signal processing flow of a photonic RF transversal signal processor with an optical microcomb source. EOM: electro-optic modulator. PD: photodetector.

In this paper, we provide a detailed analysis for both theoretical limitations and experimental factors that affect the processing accuracy of microcomb-based photonic RF transversal signal processors. First, the theoretical limitations determined by tap number, signal bandwidth, and pulse waveform are investigated. Next, the processing errors induced by imperfect responses of the practical components of the transversal signal processors are discussed, together with the methods used to reduce these errors. Finally, the theoretically calculated and experimentally measured processing errors are compared to analyze their relative contributions to the overall processing inaccuracy. These results provide a guide for the design of microcomb-based photonic RF transversal signal processors and serve as a roadmap for improving their processing accuracy.

## II. MICROCOMB-BASED PHOTONIC RF TRANSVERSAL SIGNAL PROCESSORS

RF transversal signal processors are designed based on the classical transversal filter structure that has found wide applications in signal processing [38]. They are equivalent to digital signal processors with a finite impulse response. The implementation of RF transversal signal processors based on photonic technologies and hardware can overcome the electrical bandwidth bottleneck and provide significantly increased processing bandwidth [19]. **Fig. 1** shows the schematic diagram and signal processing flow of a photonic RF transversal signal processor, where an optical microcomb generated by a compact chip-scale resonator serves as a multi-wavelength source that provides a large number of discrete wavelength channels or taps. An input RF signal is multicast onto different wavelength channels by using an electro-optic modulator (EOM), and then successive time delays between different wavelength channels are introduced by a dispersive module. Next, the delayed signal in each wavelength channel gets weighted by an optical spectral shaping module. Finally, the delayed and weighted signals are summed upon photodetection and converted back into an RF signal as the ultimate system output.

The impulse response of the photonic RF transversal signal processor in **Fig. 1** is given by [19, 20]

$$h(t) = \sum_{n=0}^{M-1} a_n \delta(t - n\Delta t), \qquad (1)$$

where $M$ is the number of taps, $a_n$ ($n = 0, 1, 2, \ldots, M\text{-}1$) is the tap weight of the $n^{th}$ tap, and $\Delta t$ is the time delay between adjacent taps. The output RF signal $s(t)$ can be expressed as [39]

$$s(t) = f(t) * h(t) = \sum_{n=0}^{M-1} a_n f(t - n\Delta t), \qquad (2)$$

where $f(t)$ is the input RF signal. After Fourier transformation from **Eq. (1)**, the spectral transfer function of the photonic RF transversal signal processor can be expressed as

$$H(\omega) = \sum_{n=0}^{M-1} a_n e^{-j\omega n\Delta t}, \qquad (3)$$

which is consistent with the spectral response of classical RF transversal filters [39].

According to **Eqs. (1) – (3)**, different signal processing functions can be realized by designing the corresponding tap weights $a_n$ ($n = 0, 1, 2, \ldots, M\text{-}1$) [19]. This forms the basis for a reconfigurable photonic RF transversal signal processor based on a single system – simply achieved by varying the comb shaping according to specific required tap coefficients, without any change in hardware.

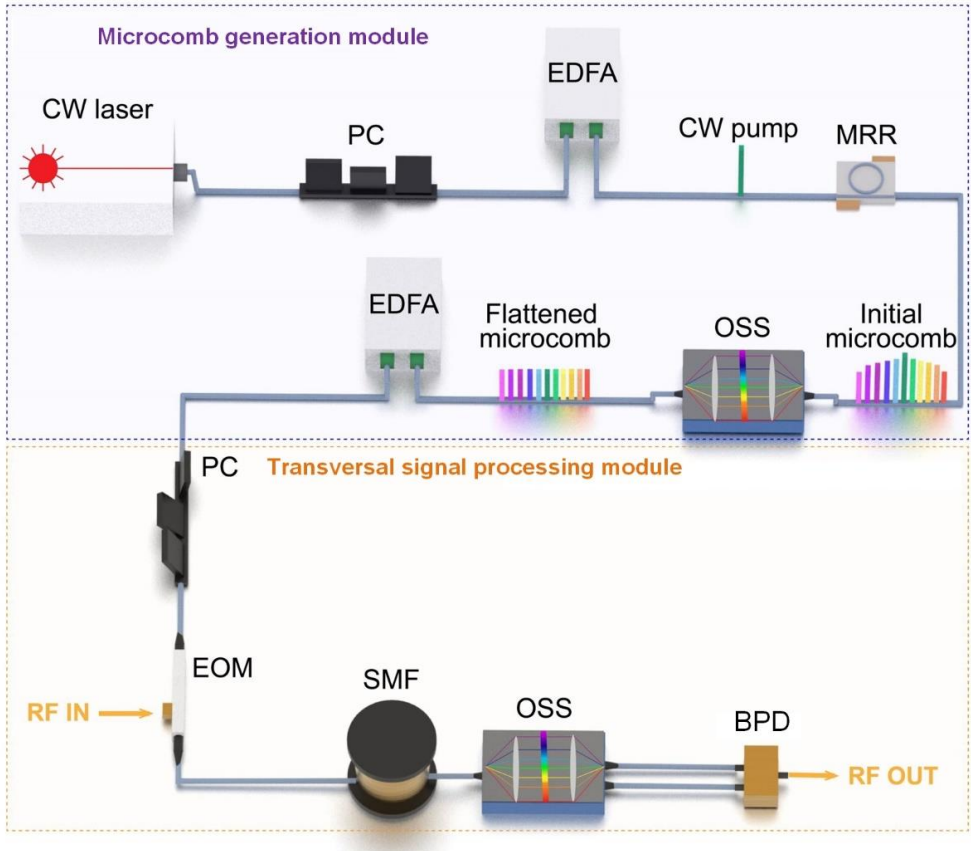

**Fig. 2.** Schematic of the practical implementation of a microcomb-based photonic RF transversal signal processor. PC: polarization controller. EDFA: erbium-doped fibre amplifier. MRR: microring resonator. OSS: optical spectral shaper. EOM: electro-optic modulator. SMF: single-mode fibre. BPD: balanced photodetector.

**Fig. 2** shows a practical implementation of a photonic RF transversal signal processor with an optical microcomb source, which includes a microcomb generation module and a transversal signal processing module. In the microcomb generation module, a continuous-wave (CW) light amplified by an erbium-doped fibre amplifier (EDFA) is employed to pump a high-Q nonlinear microring resonator (MRR) for generating initial optical microcombs, and a polarization controller (PC) is used to adjust the input polarization. The initial optical microcombs with uneven powers for different comb lines are shaped by an optical spectral shaper to flatten the microcomb lines. The flattened optical microcomb from the microcomb generation module is then sent to the transversal signal processing module that performs the signal processing flow illustrated in **Fig. 1**. The processing module consists of a PC, an EOM, a spool of single-mode fibre (SMF) as the dispersive module, an optical spectral shaper (OSS) as the spectral shaping module, and a balanced photodetector (BPD) for photodetection. BPD connecting the two complementary output ports of the OSS is used to divide all the wavelength channels into two groups and introduce a phase difference of $\pi$ between them, thus allowing for both positive and negative signs of the tap coefficients.

Processing accuracy is a key parameter for characterizing the performance of signal processors. For microcomb-based photonic RF transversal signal processors, the deviations between the output of the signal processors and the ideal signal processing results arise not only from the theoretical approximation of continuous impulse response (which corresponds to infinite tap number $M$) using practical systems with finite tap numbers, but also error sources from different components of practical systems. The former results in the variation of the processing accuracy with tap number, signal bandwidth, and pulse waveform. The latter mainly includes phase noise of microcombs, uneven gain and noise of EDFA, chirp and limited modulation bandwidth of EOM, phase errors induced by SMF, shaping errors induced by OSSs, and noise and uneven transmission response of BPD. In the following sections, the processing accuracy of microcomb-based photonic RF transversal signal processors is analyzed and discussed in detail. In Section III, we investigate the theoretical limitations of the processing accuracy determined by tap number, signal bandwidth, and pulse waveform. In section IV, we provide a discussion of the error sources in practical systems. In section V, we analyze the contributions of the theoretical limitations and the experimental factors to the overall processing inaccuracy.

## III. THEORETICAL LIMITATIONS OF PROCESSING ACCURACY

From **Eq. (1)**, it can be seen that the delayed taps in a photonic RF transversal signal processor can be considered as discrete samples of the input RF signal, and the processing output is the sum of the weighted discrete samples. In principle, for a tap number approaching infinity ($M \to \infty$), any arbitrary impulse response for different signal processing functions can be realized by applying corresponding tap coefficients to each wavelength channel. However, for a practical system with a finite $M$, there are deviations between its processing output and the ideal results that decrease with increasing $M$. At the same time, the sampling rate determines the free spectral range (FSR) of the RF spectral response of the transversal signal processor, which is given by [20]

$$FSR_{RF} = \frac{1}{\Delta t},\qquad(4)$$

where $\Delta t$ is the time delay between adjacent taps that can be expressed as [20]

$$\Delta t = L \cdot D \cdot \Delta\lambda,\qquad(5)$$

where $L$ is the length of the dispersive medium, $D$ is the dispersion parameter, and $\Delta\lambda$ is the comb spacing. According to **Eqs. (4) - (5)**, $FSR_{RF}$ can be varied by changing $L$, $D$, or $\Delta\lambda$, and different $FSR_{RF}$ results in the differences between the processing accuracy for input signals with different spectral bandwidths and pulse waveforms (*i.e.*, frequency components).

In this section, we use three typical signal processing functions including differentiation, integration, and the Hilbert transform as examples to investigate the theoretical limitations in the processing accuracy of microcomb-based photonic RF transversal signal processors determined by tap number, signal bandwidth, and pulse waveform. The transfer function of $N$th-order temporal differentiation can be expressed as [16]

$$H_{diff}(\omega) = (j\omega)^N,\qquad(6)$$

where $j = \sqrt{-1}$, $\omega$ is the angular frequency, and $N$ is the order of the differentiator that can be either an integer or a fraction. The $N$th-order integration is the inversion of the $N$th-order differentiation, and the transfer function can be given by [17]

$$H_{int}(\omega) = \left(\frac{1}{j\omega}\right)^N,\qquad(7)$$

The transfer function of an integral or fractional Hilbert transformer can be expressed as [18]

$$H(\omega) = \begin{cases} e^{-j\phi}, & 0 \le \omega < \pi \\ e^{j\phi}, & -\pi \le \omega < 0 \end{cases}\qquad(8)$$

where $\phi = N \times \pi/2$ is the phase shift, with $N$ denoting the order of the Hilbert transform, which can be either 1 or a fraction.

To quantify the comparison of processing accuracy, the root mean square error (RMSE) is introduced in our analysis, which is defined as [40]

$$RMSE = \sqrt{\sum_{i=1}^{k}\frac{(Y_i \cdot y_i)^2}{k}}\qquad(9)$$

where $k$ is the number of sampled points, $Y_1$, $Y_2$, …, $Y_n$ are the values of the ideal signal processing result, and $y_1$, $y_2$, …, $y_n$ are the values of the simulated output of the microcomb-based photonic RF transversal signal processors. For the analysis in this section, we only consider the processing inaccuracy induced by the processor's limited tap number and the input RF signal's spectral bandwidth and temporal waveform. We do not account for experimental processing errors induced by the imperfect response of components in practical systems – this is discussed in Section IV.

### A. Influence of tap number

Although there are discrepancies between a processor's output and the ideal results for practical systems having a limited number of taps $M$, this can be made negligible as $M$ becomes sufficiently large. Here we analyze the influence of the tap number on the processing accuracy of microcomb-based photonic RF transversal signal processors.

**Fig. 3(a)** shows the output waveforms of microcomb-based photonic RF transversal signal processors that perform $N$th-order integral differentiation ($N = 1, 2, 3, 5, 8$, and 10 in **Eq. (6)**), together with the ideal results. The input RF signal is a Gaussian pulse with a full-width-at-half-maximum (FWHM) of ~0.17 ns. The parameters in **Eq. (5)** of the transversal signal processor are assumed to be $\Delta\lambda = 0.4$ nm, $L = 4.8$ km, and $D = 17.4$ ps/nm/km, which results in an $FSR_{RF}$ of ~30 GHz (*i.e.*, $\Delta t = $ ~0.033 ns). As the tap number $M$ increases from 10 to 160, the processors' output waveforms match the ideal results better for all the integral differentiation orders, reflecting the improved processing accuracy that can be achieved by increasing the tap number.

**Fig. 3(b)** shows RMSEs between the processors' output waveforms and the ideal integral differentiation as a function of $M$. As expected, the RMSEs decrease with increasing $M$ for all $N$, showing agreement with the trend in **Fig. 3(a)**. For smaller tap numbers, the RMSEs decrease more sharply, especially for $M \le 30$ and higher-order differentiation. As the tap number increases, the decrease in RMSEs becomes more gradual. For $M \ge 80$, there is only a very small improvement in the processing accuracy with increasing $M$. This indicates that the need for very large tap numbers (*e.g.*, $M > 100$) in practical applications is not strong, particularly given the limited operation bandwidth arising from the required reduced comb spacing and increased complexity of comb shaping. We also note that the RMSEs increase as the integral differentiation order $N$ increases from 1 to 10. Such an increase becomes less obvious as the tap number $M$ increases, and for $M \ge 80$, there is only a very small difference between the RMSEs for different $N$. This reflects the high processing accuracy for high-order differentiation when $M$ is sufficiently high, which is another important advantage of microcomb-based photonic RF transversal signal processors. In contrast, high-order photonic differentiators based on multiple cascaded passive resonators [41, 42] suffer from limitations induced by misalignment of resonance wavelengths from separate subunits, with their processing accuracy degrading rapidly with increasing differentiation order.

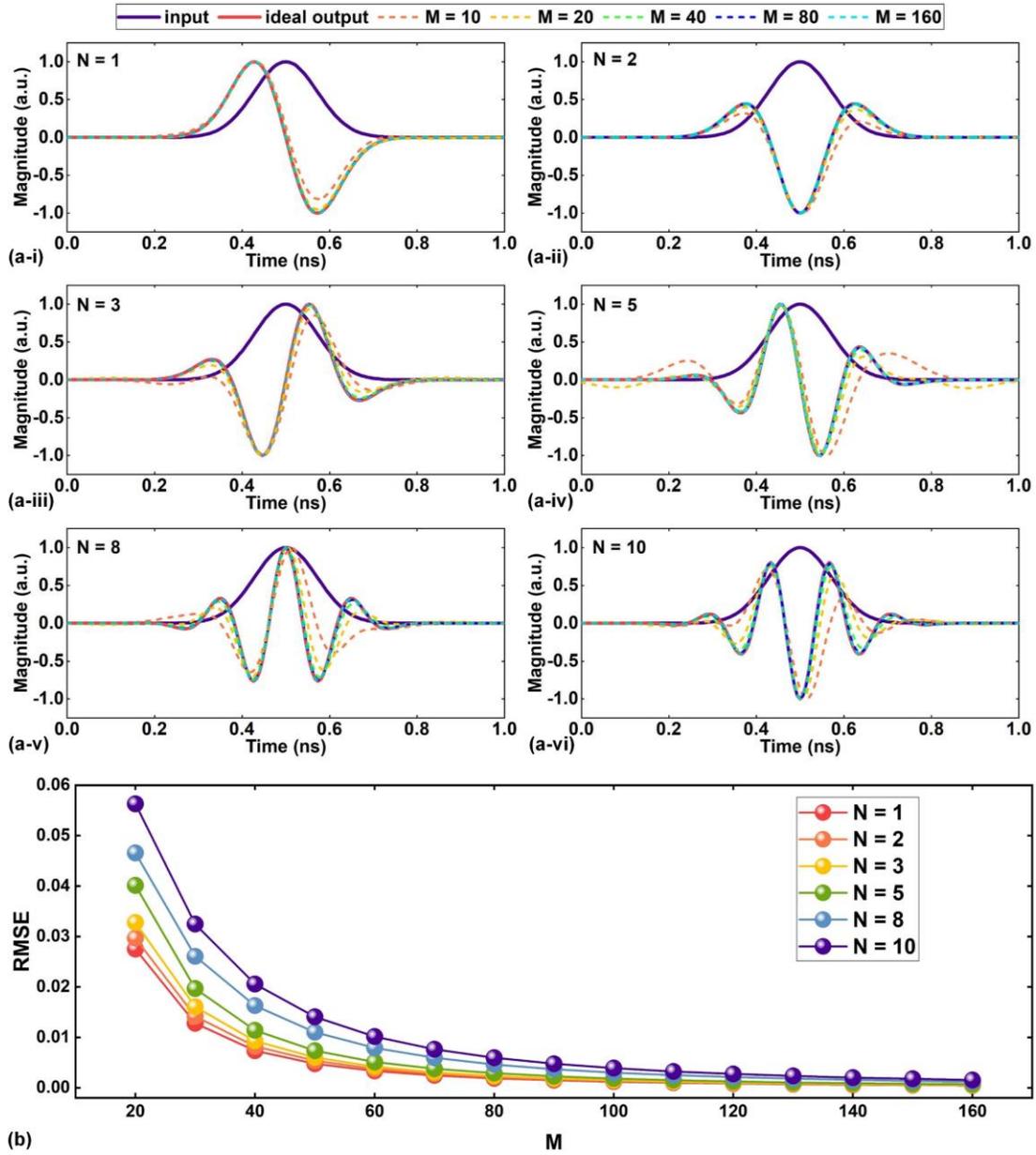

**Fig. 3.** Influence of tap number on the processing accuracy of microcomb-based photonic RF transversal signal processors that perform $N^{th}$-order integral differentiation ($N$ = 1, 2, 3, 5, 8, 10). (a) Temporal waveform of Gaussian input pulse and output waveforms from the processors with tap numbers $M$ = 10 – 160. The ideal differentiation results are also shown for comparison. (b) Root mean square errors (RMSEs) between the ideal differentiation results and the processors' output waveforms as a function of $M$. In (a) and (b), the Gaussian input pulse has a full width at half maximum (FWHM) of ~0.17 ns. The comb spacing, length of dispersive medium, and dispersion parameter are $\Delta\lambda$ = 0.4 nm, $L$ = 4.8 km, and $D$ = 17.4 ps/nm/km, respectively, which allow for an $FSR_{RF}$ of ~30 GHz.

**Fig. 4(a)** shows the output waveforms from microcomb-based photonic RF transversal signal processors that perform $N^{th}$-order fractional differentiation ($N$ = 0.15, 0.3, 0.45, 0.6, 0.75, and 0.9 in **Eq. (6)**), together with the ideal results. The input RF signal and the parameters of the microcomb and the dispersive medium are kept the same as those in **Fig. 3**. For all the fractional differentiation orders, the processors' output waveforms match the ideal results better as the tap number $M$ increases from 10 to 160, showing a trend similar to that in **Fig. 3**.

**Fig. 4(b)** shows RMSEs between the processors' output waveforms and the ideal fractional differentiation results as a function of $M$. The RMSEs decrease with $M$ for all $N$, showing agreement with the trend in **Fig. 4(a)**. Compared with **Fig. 3(b)**, both the RMSEs and their rates of decrease with $M$ are much lower, mainly because the roll-off in frequency response of the integral differentiators with $N$ >1 is much steeper than the fractional differentiators with $N$ < 1. This also reflects that the need for large tap numbers for fractional differentiators is even lower.

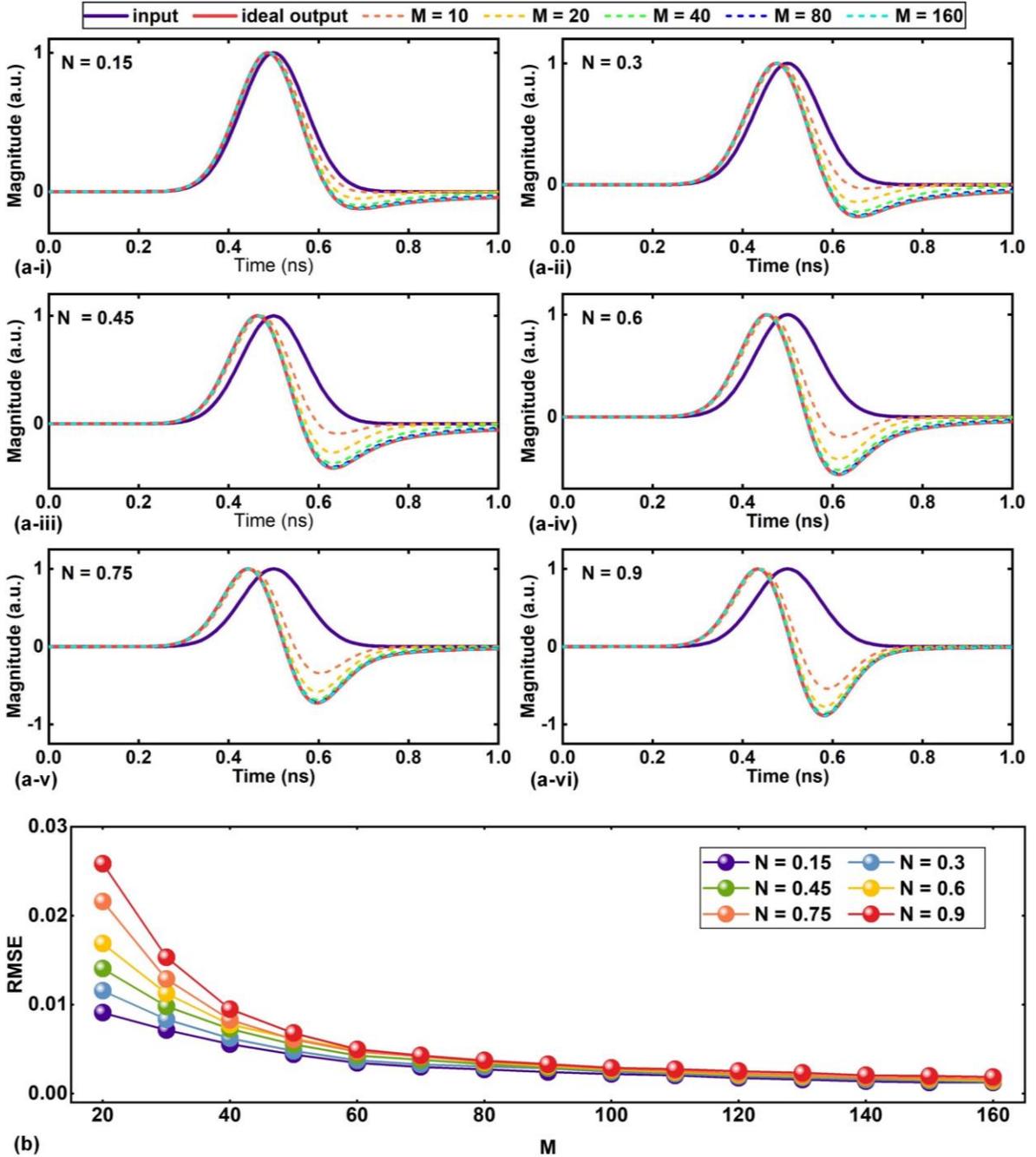

**Fig. 4.** Influence of tap number on the processing accuracy of microcomb-based photonic RF transversal signal processors that perform $N$th-order fractional differentiation ($N$ = 0.15, 0.3, 0.45, 0.6, 0.75, 0.9). (a) Temporal waveform of Gaussian input pulse and output waveforms from the processors with tap numbers $M$ = 10 − 160. The ideal differentiation results are also shown for comparison. (b) RMSEs between the ideal differentiation results and the processors' output waveforms as a function of $M$. In (a) and (b), the Gaussian input pulse has a full width at half maximum (FWHM) of ~0.17 ns. The comb spacing, length of dispersive medium, and dispersion parameter are $\Delta\lambda$ = 0.4 nm, $L$ = 4.8 km, and $D$ = 17.4 ps/nm/km, respectively, which allow for an $FSR_{RF}$ of ~30 GHz.

Fig. 5(a) shows the output waveforms from microcomb-based photonic RF transversal signal processors that perform 1st-order integration ($N$ = 1 in **Eq. (7)**), together with the ideal results. The input RF signal and the parameters of the microcomb and the dispersive medium are kept the same as those in **Fig. 3**. In **Figs. 5(a-i)**, we show the integration results for a single Gaussian input pulse. In **Figs. 5(a-ii)** – **(a-iv)**, we show the integration results for dual Gaussian input pulses with different time intervals of 0.2, 0.4, and 1.0 ns, respectively. For all the integration results in **Fig. 5(a)**, the processors' output waveforms agree better with the ideal results as the tap number $M$ increases. This is because the length of the integration time window $T$ (defined as $T = M \times L \times D \times \Delta\lambda$) increases proportionally with $M$, and the increase of $M$ yields longer integration windows that are closer to the infinite integration window length for ideal integration.

Fig. 5(b) shows RMSEs between the processors' output waveforms and the ideal integration results as a function of $M$. Since the integration window varies for different $M$, for a fair comparison, here we choose a time window start at 0 s and end when ideal results reach the amplitude of 1 to calculate the RMSEs. The calculated RMSEs decrease with $M$ for all different Gaussian input RF signals, which agrees with the trend in **Fig. 5(a)**. For dual Gaussian input pulses, especially those with a large time interval, RMSEs are relatively large when $M$ < 60, mainly resulting from the fact that the narrow integration time window $T$ for small taps (*e.g.*, $T$ = ~0.65 ns for $M$ = 20) cannot cover the time intervals between the dual Gaussian input pulses. This indicates that for an input signal with a long time duration, a large tap number is needed to provide a wide integration window.

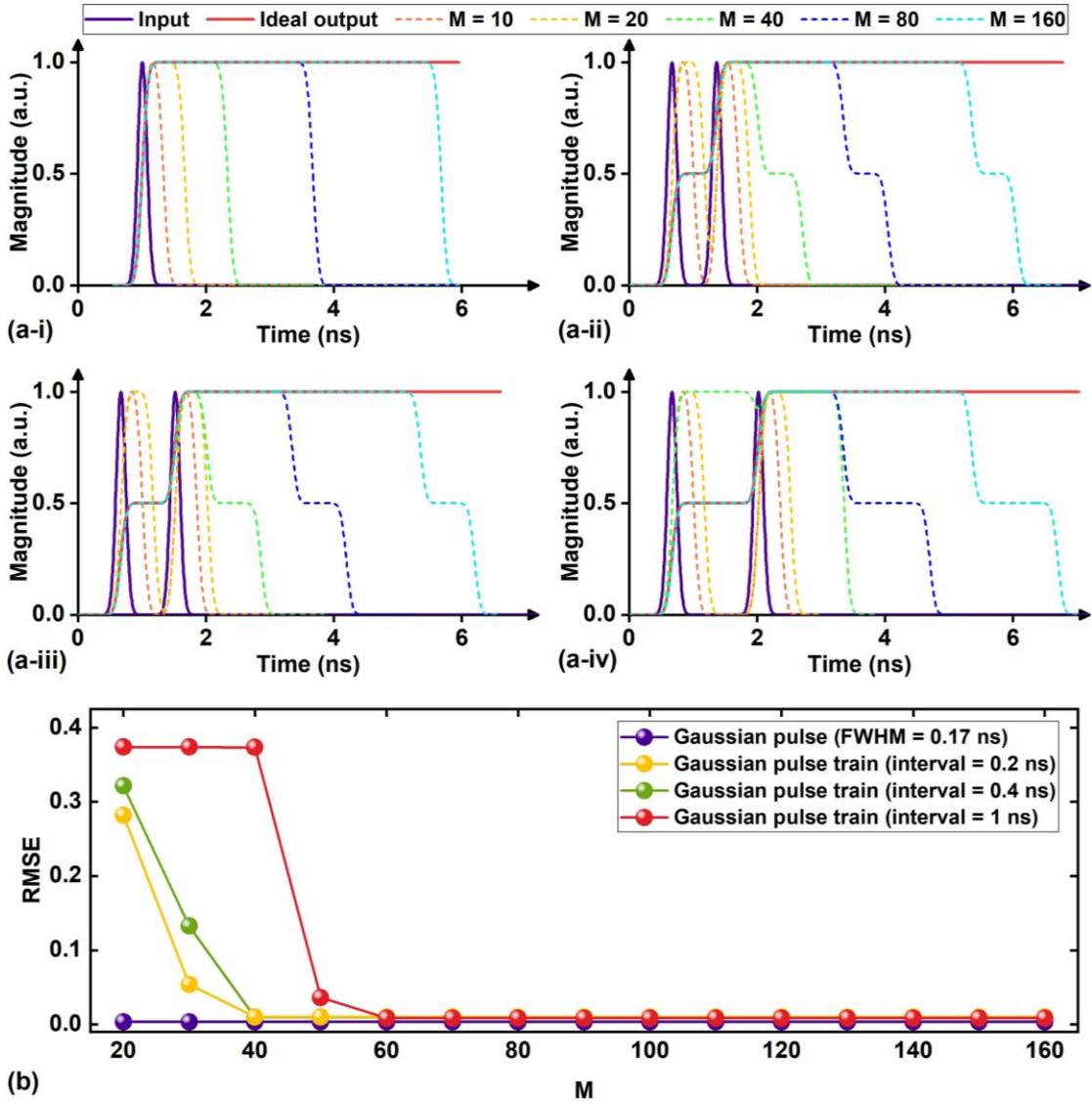

**Fig. 5.** Influence of tap number on the processing accuracy of microcomb-based photonic RF transversal signal processors that perform 1st-order integration ($N$ = 1). (a) Temporal waveform of Gaussian input pulse and output waveforms from the processors with tap numbers $M$ = 10 − 160. The ideal integration results are also shown for comparison. (b) RMSEs between the ideal integration results and the processors' output waveforms as a function of $M$. In (a), the FWHMs of the Gaussian input pulses are ~0.17 ns. The time intervals of the dual Gaussian pulses in (ii) – (iv) are ~0.2, ~0.4, and ~1.0 ns, respectively. In (a) and (b), the comb spacing, length of dispersive medium, and dispersion parameter are $\Delta\lambda$ = 0.4 nm, $L$ = 4.8 km, and $D$ = 17.4 ps/nm/km, respectively, which allow for an $FSR_{RF}$ of ~30 GHz.

**Fig. 6(a)** shows the output waveforms from microcomb-based photonic RF transversal signal processors that perform Hilbert transforms with a phase shift of $\phi$ = 15°, 30°, 45°, 60°, 75°, and 90° in **Eq. (8)**, which correspond to Hilbert transform orders $N$ = 0.166, 0.333, 0.5, 0.667, 0.833, and 1, respectively.

The ideal results for the Hilbert transforms are also shown. The input RF signal and the parameters of the microcomb and the dispersive medium are kept the same as those in **Fig. 3**. Similar to those in **Figs. 3 − 5**, the processors' output waveforms match the ideal results better for all $\phi$ as the tap number $M$ increases.

**Fig. 6(b)** shows RMSEs between the processors' output waveforms and the ideal Hilbert transform results as a function of $M$. The RMSEs decrease with $M$ for all $\phi$, agreeing with the trend in **Fig. 6(a)**. The RMSEs increase as the phase shift $\phi$ increases from 15° to 90°, mainly due to the fact that the amplitude ripple within the passband of the frequency response is more significant for a larger phase shift.

Although the processing accuracy increases with tap number $M$ for all signal processing functions in **Figs. 3 − 6**, it should be noted that increasing the tap number can also limit practical microcomb-based photonic RF transversal signal processors. There are limited operation bandwidths for many of the components in the transversal signal processors, for example, the EDFA and EOM in **Fig. 2** have typical operation bandwidths in the telecom C-band (from 1530 to 1565 nm). This results in limited operation bandwidths for practical transversal signal processors, where increased tap numbers can only be achieved by reducing the comb spacing $\Delta\lambda$. On the other hand, to avoid the overlap between the modulated RF replicas on different wavelength channels, the comb spacing needs to be at least twice the bandwidth of the input RF signal, and so a narrow comb spacing would limit the bandwidth of the RF signal to be processed. In practical applications, one needs to properly balance the trade-off between tap number and comb spacing. In **Figs. 3 − 6**, there is only a very small improvement in the processing accuracy for $M \geq 80$, therefore a tap number $M$ = 80 has been widely used for practical microcomb-based photonic RF transversal signal processors operating in the C-band [17, 18, 31-33], which corresponds to a comb spacing of ~0.4 nm (*i.e.*, ~50 GHz).

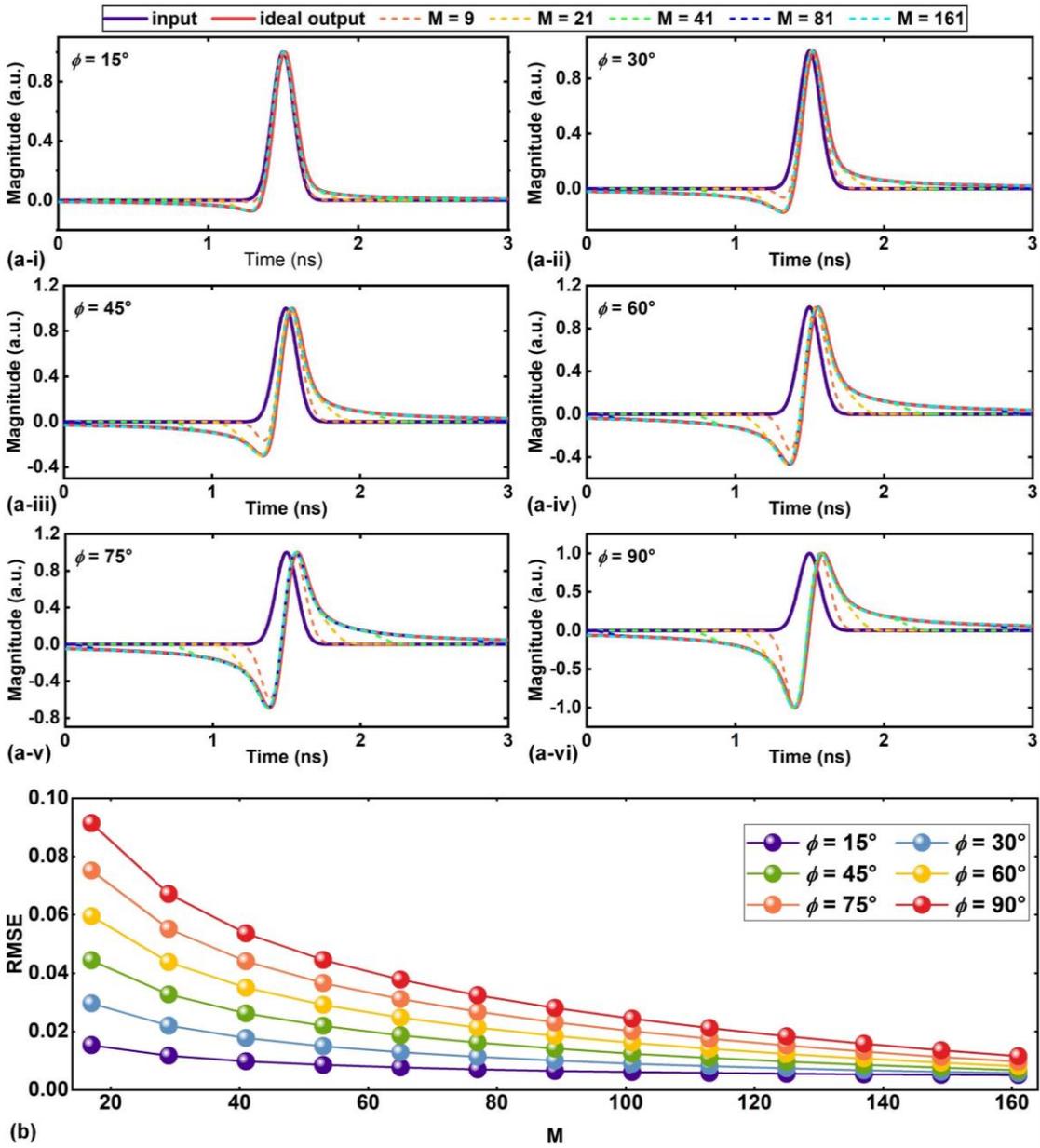

**Fig. 6.** Influence of tap number on the processing accuracy of microcomb-based photonic RF transversal signal processors that perform Hilbert transform with a phase shift of $\phi$ = 15°, 30°, 45°, 60°, 75°, and 90°. (a) Temporal waveform of Gaussian input pulse and output waveforms from the processors with tap numbers $M$ = 9 − 161. The ideal Hilbert transform results are also shown for comparison. (b) RMSEs between the ideal Hilbert transform results and the processors' output waveforms as a function of $M$. In (a) and (b), the Gaussian input pulse has a full width at half maximum (FWHM) of ~0.17 ns. The comb spacing, length of dispersive medium, and dispersion parameter are $\Delta\lambda$ = 0.4 nm, $L$ = 4.8 km, and $D$ = 17.4 ps/nm/km, respectively, which allow for an $FSR_{RF}$ of ~30 GHz.

### B. Influence of signal bandwidth

The processing accuracy will also be affected by the bandwidth of the input RF signal. According to the Nyquist sampling theorem, a band-limited continuous-time signal needs to be sampled more than twice as fast as its highest frequency component to avoid overlap [43]. This sets a lower limit for the $FSR_{RF}$ in **Eq. (4)**, which needs to be at least twice as large as the bandwidth of the input RF signal. On the other hand, the influence of the input signal's bandwidth on the processing accuracy also varies for different signal processing functions due to the differences in their frequency response. Here, we analyze the influence of the input signal bandwidth on the processing accuracy of differentiators, integrators, and Hilbert transformers.

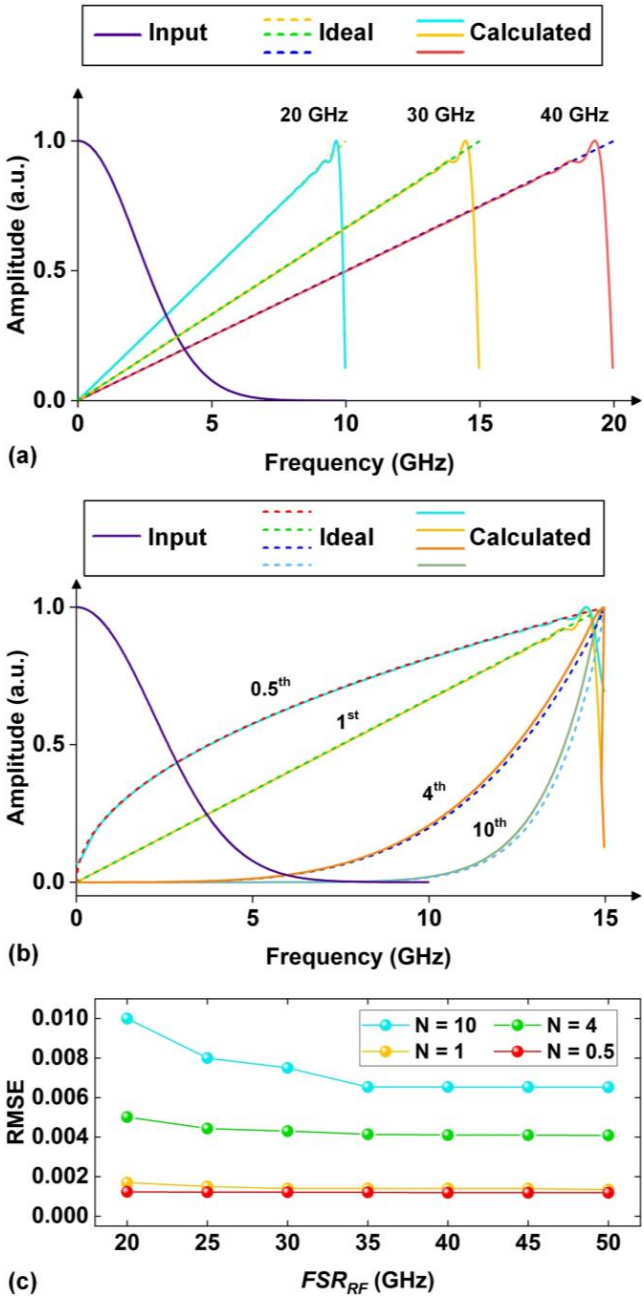

**Fig. 7.** Influence of input signal bandwidth on the processing accuracy of microcomb-based photonic RF transversal signal processors that perform differentiations. (a) Spectrum of Gaussian input RF signal and the amplitude frequency response of the processors that perform 1-st order differentiation with $FSR_{RF}$ = 20, 30, and 40 GHz. The ideal response is also shown for comparison. (b) Amplitude frequency response of the processors that perform $N^{th}$-order differentiation ($N$ = 0.5, 1, 4, and 10) for fixed $FSR_{RF}$ = 30 GHz. The spectrum of the Gaussian input RF signal same as that in (a) and the ideal response are also shown for comparison. (c) RMSEs between the ideal differentiation results and the processors' output waveforms as a function of $FSR_{RF}$ for different $N$ = 0.5, 1, 4, and 10. In (a) – (c), the Gaussian input pulse has a FWHM of ~0.17 ns. The tap number and comb spacing are $M$ = 80 and $\Delta\lambda$ = 0.4 nm, respectively.

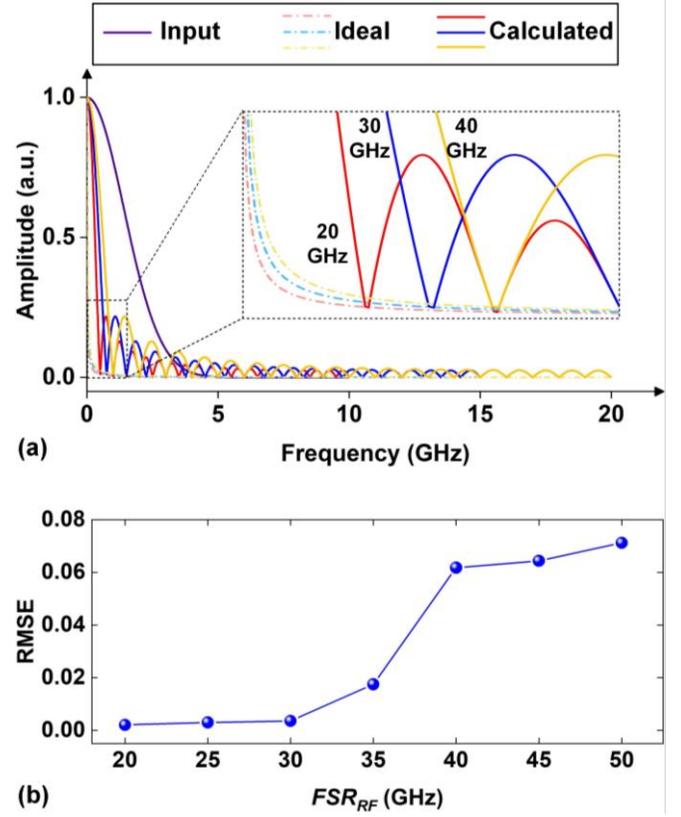

**Fig. 8.** Influence of input signal bandwidth on the processing accuracy of microcomb-based photonic RF transversal signal processors that perform 1st-order integration. (a) Spectrum of Gaussian input RF signal and the amplitude frequency response of the processors with $FSR_{RF}$ = 20, 30, and 40 GHz. The ideal response is also shown for comparison. (b) RMSEs between the ideal integration results and the processors' output waveforms as a function of $FSR_{RF}$. In (a) and (b), the Gaussian input pulse has a FWHM of ~0.17 ns. The tap number and comb spacing are $M$ = 80 and $\Delta\lambda$ = 0.4 nm, respectively.

**Fig. 7(a)** shows the spectrum of a Gaussian input RF signal with a FWHM of ~0.17 ns and the amplitude frequency response of a microcomb-based photonic RF transversal signal processor that performs 1$^{st}$-order differentiation ($N$ = 1). We plot the frequency response for different $FSR_{RF}$ of 20, 30, and 40 GHz, together with the ideal amplitude response for each $FSR_{RF}$. For all $FSR_{RF}$, the deviation between the actual and ideal amplitude response is larger at high frequency region. Further, a larger $FSR_{RF}$ yields a better overlap between the input signal bandwidth and the low-error region of the amplitude response, resulting in better agreement and improved processing accuracy.

**Fig. 7(b)** shows the amplitude frequency response of microcomb-based photonic RF transversal signal processors for different differentiation orders $N$ = 0.5, 1, 4, and 10, together with the ideal amplitude response for each $N$. The Gaussian input RF signal is the same as in **Fig. 7(a)**, and the $FSR_{RF}$ of the processors is kept the same as 30 GHz. For a higher differentiation order $N$, the deviation in the amplitude response is more significant, particularly at high frequencies. This indicates that high-order differentiators need a larger $FSR_{RF}$ to achieve the same processing accuracy as low-order differentiators.

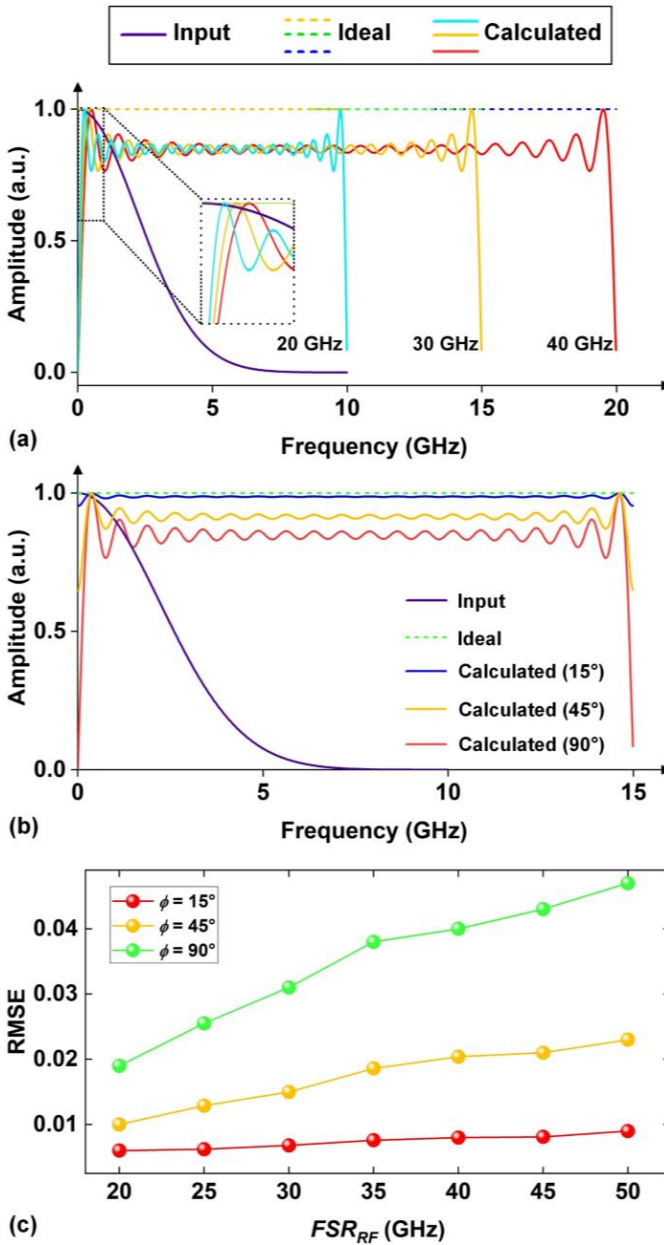

**Fig. 9.** Influence of input signal bandwidth on the processing accuracy of microcomb-based photonic RF transversal signal processors that perform Hilbert transform with a phase shift of $\phi$ = 15°, 45°, and 90°. (a) Spectrum of Gaussian input RF signal and the amplitude frequency response of the processors with $FSR_{RF}$ = 20, 30, and 40 GHz. The ideal response is also shown for comparison. (b) RMSEs between the ideal integration results and the processors' output waveforms as a function of $FSR_{RF}$. In (a) and (b), the Gaussian input pulse has a FWHM of ~0.17 ns. The tap number and comb spacing are $M$ = 81 and $\Delta\lambda$ = 0.4 nm, respectively.

**Fig. 7(c)** shows the RMSEs between the processors' output waveforms and the ideal differentiation results versus $FSR_{RF}$ for different $N$. As expected, the RMSEs decrease with $FSR_{RF}$, agreeing with the trend in **Fig. 7(a)**. On the other hand, larger RMSEs are produced for larger $N$, although also decrease more sharply with $FSR_{RF}$. This is consistent with the trend in **Fig. 7(b)**. The decrease in RMSEs becomes more gradual as $FSR_{RF}$ increases, being negligible for large $FSR_{RF}$. This is because for large $FSR_{RF}$ the input RF signal spectrum overlaps better with the low-error region of the amplitude response.

**Fig. 8** shows the analysis of the influence of signal bandwidth on the processing accuracy for 1st-order integration. The input RF signal and transversal signal processor are the same as those in **Fig. 7**. For all $FSR_{RF}$, the deviation in the processor's amplitude response is larger at low frequency region, resulting in improved processing accuracy for a smaller $FSR_{RF}$. This is opposite to the trend for differentiation in **Fig. 7**, resulting mainly from the difference in their frequency response. Although a smaller $FSR_{RF}$ yields better processing accuracy, $FSR_{RF}$ still needs to be larger than the lower limit set by the Nyquist sampling theorem (*i.e.*, twice the input signal bandwidth) to avoid overlap of the modulated sidebands.

**Fig. 9** shows the corresponding results for analysing the influence of the signal bandwidth on the processing accuracy of Hilbert transformers. The input RF signal and transversal signal processor are the same as **Figs. 7** and **8**. For all $FSR_{RF}$, the RMSEs increase with $FSR_{RF}$, showing a trend similar to integration in **Fig. 8** but opposite to differentiation in **Fig. 7**. This is because the deviation in the processor's amplitude response is mainly induced by the amplitude ripples at low frequencies and a smaller $FSR_{RF}$ results in a lower bandwidth of the ripples that make them overlap less with the high-intensity frequency components of the input RF signal.

According to **Figs. 7 – 9**, in order to optimize the processing accuracy of a microcomb-based photonic RF transversal signal processor, the $FSR_{RF}$ needs to be tailored according to the bandwidth of the input RF signal and the specific processing function. According to **Eqs. (4)** and **(5)**, for fixed comb spacing and tap number, the $FSR_{RF}$ can be changed by varying the length $L$ or dispersion parameter $D$ of the dispersive medium. It should also be noted that although in principle very large $FSR_{RF}$ can be achieved by choosing a very small $L$ or $D$, the operation bandwidth for practical transversal signal processors is still limited by the maximum Nyquist zone according to the Nyquist sampling theorem, which is half of the comb spacing $\Delta\lambda$.

### C. Influence of signal waveform

As can be seen from the analysis in the previous subsection, the processing accuracy is affected by the overlap between the input RF signal spectrum and the response spectrum of the transversal signal processor. Therefore, the processor's output error will vary for different input RF signal waveforms and spectral profiles. Here, we analyze the influence of the input waveform on the accuracy of different processing functions.

We investigate four different input RF signals, including Gaussian ($m$ = 1), super Gaussian ($m$ = 3), triangle, and parabolic pulses. The temporal waveforms of Gaussian and super Gaussian pulses can be expressed as [44]:

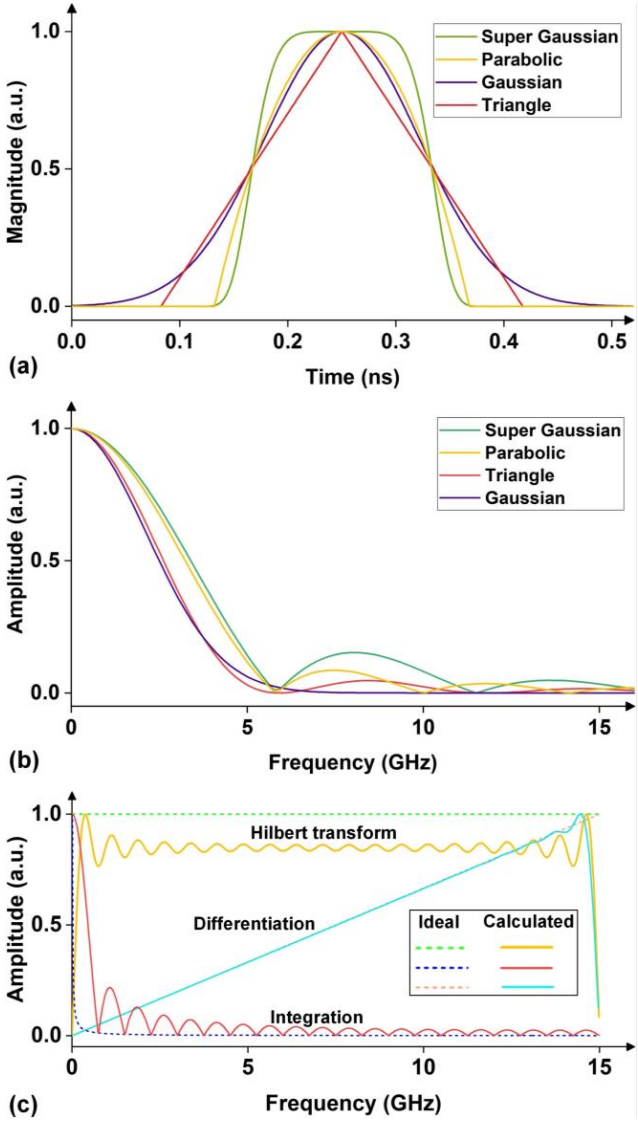

**Fig. 10.** (a) Temporal waveforms and (b) corresponding amplitude spectra of different input RF signals including Gaussian ($m = 1$), super Gaussian ($m = 3$), triangle, and parabolic pulses. All the RF signals have the same FWHM of ~0.17 ns. (c) Amplitude frequency response of microcomb-based photonic RF transversal signal processors that perform 1-st order differentiation, integration, and Hilbert transform. All the processors have the same parameters of $FSR_{RF} = 30$ GHz and $\Delta\lambda = 0.4$ nm. The tap numbers for differentiation, integration, and Hilbert transform are $M = 80$, 80, and 81, respectively.

$$x_g(t) = \sqrt{P} \cdot exp\left[-\frac{1}{2}\left(\frac{t}{\tau}\right)^{2 \cdot m}\right], \qquad (10)$$

where $P$ is the pulse peak power, $\tau$ is the half-width at $1/e$ intensity, and $m$ is the Gaussian pulse order, with $m = 1$ and $m > 1$ corresponding to the Gaussian and super Gaussian pulses, respectively. The temporal waveform of a triangle pulse is given by [45]

$$x_t(t) = \frac{4P}{T}\left|\left[\left(t - \frac{T}{4}\right) \cdot \text{MOD } T\right] - \frac{T}{2}\right| - P, \qquad (11)$$

where MOD denotes the modulo operation, $P$ and $T$ are the amplitude and period, respectively. The temporal waveform of a parabolic pulse can be expressed as [46]

$$x_p(t) = \begin{cases} P\left(1 - \left(\frac{t}{T/2}\right)^2\right), & |t| < T/2 \\ 0, & |t| \geq T/2 \end{cases}, \qquad (12)$$

where $P$ and $T$ are the pulse peak power and period, respectively.

**Fig. 10(a)** shows the temporal waveforms of Gaussian ($m = 1$), super Gaussian ($m = 3$), triangle, and parabolic pulses. For comparison, all signals have the same FWHM of ~0.17 ns. The amplitude spectra of the signals in **Fig. 10(a)** are shown in **Fig. 10(b)**. As can be seen, different temporal waveforms yield different intensity spectra. For example, the parabolic and super Gaussian pulses have stronger high frequency components than the other two pulse waveforms. **Fig. 10(c)** shows the amplitude frequency response of microcomb-based photonic RF transversal signal processors that perform 1st order differentiation, integration, and Hilbert transform (*i.e.*, $N = 1$ in **Eqs. (6) − (8)**). All the three processors have the same $FSR_{RF} = 30$ GHz and $\Delta\lambda = 0.4$ nm, and the tap numbers for differentiation, integration, and Hilbert transform are $M = 80$, 80, and 81, respectively. As expected, an improved processing accuracy can be achieved when there is better overlap between the high-intensity frequency components of the input RF signal and the low-error region of the processor's response spectrum.

**Figs. 11(a) − (c)** shows the output waveforms from the processors in **Fig. 10(c)** for different input RF signals in **Fig. 10(a)**, together with the ideal results. The RMSEs between the processors' output waveforms and the ideal results are shown in **Fig. 11(d)**. For differentiator, the processing accuracy of parabolic and super Gaussian pulses is lower than the other two pulse waveforms, mainly because they have stronger high frequency components that overlap with the high-error region of the processors' response spectra. Whereas for integration and Hilbert transform, the processing accuracy of parabolic and super Gaussian pulses is slightly higher than the other two waveforms since the processor's high-error region is at low frequencies. We also note that the differences in RMSEs of the integration results are not significant because the high-error region of the processor's response is within 2 GHz (as shown in **Fig. 10(c)**) whereas the differences in the input RF spectra mainly appear above 2 GHz.

## IV. ERROR SOURCES IN PRACTICAL SYSTEMS

In addition to the theoretical limitations mentioned in Section III, the imperfect response of practical components can lead to errors of microcomb-based photonic RF transversal signal processors. In this section, we discuss different error sources in practical systems, which are summarized in **Fig. 12**.

In the microcomb generation module, an important source of error is the phase noise of the microcombs. Optical microcombs with high coherence and stable mode locking are needed to achieve a high accuracy for microcomb-based photonic RF transversal signal processors over long time periods. Having said this, they can still operate with relatively incoherent microcombs since microcombs only serve as a multi-wavelength source and the optical power of the wavelength channels is incoherently detected by a PD. Regardless, however, minimizing the phase noise in the soliton microcombs improves the system performance and a variety of driving mechanisms have been proposed to achieve high microcomb coherence and stability. This includes frequency scanning [47-50], forward and backward tuning [51], two-colour pumping [52-54], EO modulation [55-57], power kicking [58-60], self-injection locking [61-63], filter-driven FWM [64-66], integrated heaters [67-69], self-referencing [70-72], cryogenic cooling [73], and piezoelectric control [74]. Recently, achieving simple mode-locking without any complex startup procedures has been reported, highlighted by demonstrations of turnkey soliton microcomb generation [75] and spontaneous generation of robust microcavity solitons [66]. In addition to the phase noise of the microcombs, the uneven gain and noise of the EDFA as well as the shaping inaccuracy caused by the OSS can also lead to amplitude and phase errors that degrade the processing accuracy.

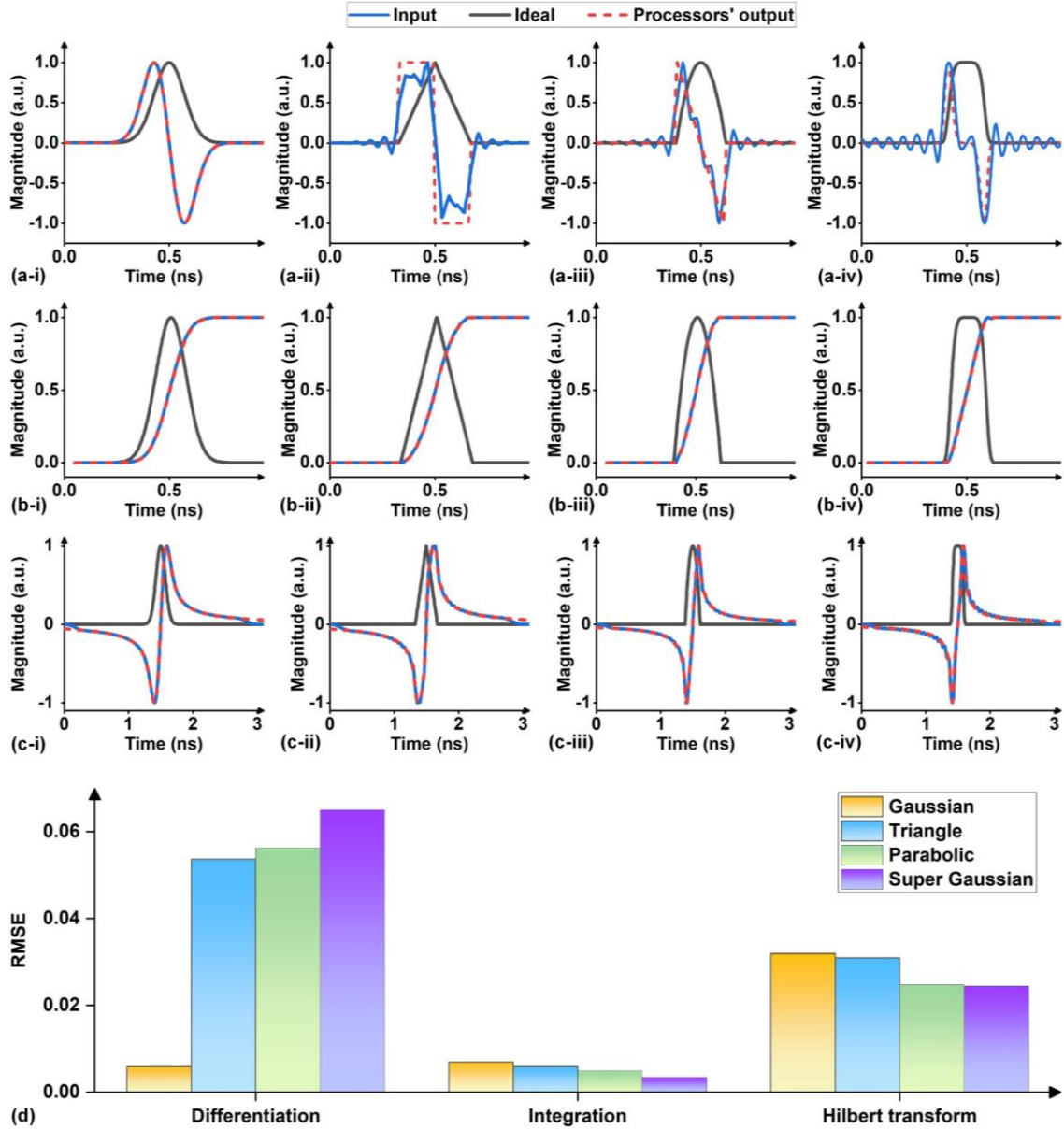

**Fig. 11.** Influence of input signal waveforms on the processing accuracy of microcomb-based photonic RF transversal signal processors that perform 1-st order differentiation, integration, and Hilbert transform. (a) – (c) Temporal waveforms of different input RF signals including (i) Gaussian ($m = 1$), (ii) triangle, (iii) parabolic, and (iv) super Gaussian ($m = 3$) pulses and the corresponding output waveforms from the processors performing differentiation, integration, and Hilbert transform, respectively. The ideal processing results are also shown for comparison. (d) RMSEs between the ideal processing results and the processors' output waveforms for different processing functions and input signal waveforms in (a) – (c). All the RF signals have the same FWHM of ~0.17 ns. All the processors have the same parameters of $FSR_{RF} = 30$ GHz and $\Delta\lambda = 0.4$ nm. The tap numbers for differentiation, integration, and Hilbert transform are $M = 80, 80,$ and 81, respectively.

In the transversal signal processing module, processing errors mainly result from imperfect amplitude and phase response of the EOM, SMF, OSS, and BPD components. For the EOM, the chirp and limited modulation bandwidth affect the processing accuracy by introducing undesired amplitude and phase errors of the modulated replicas in different wavelength channels. For the SMF that serves as the dispersive module, phase errors are introduced due to its chromatic dispersion, including both second and higher-order dispersion. For intensity modulation, typically achieved with a Mach–Zehnder modulator (MZM), double-sideband (DSB) signals with an optical carrier and two modulated sidebands are generated [76]. After photodetection, each sideband beats with the optical carrier, generating two RF beat signals that constructively interfere to produce a single composite RF output. When the modulated DSB optical signal is transmitted over fibre, the different frequency components experience different phase shifts due to chromatic dispersion, resulting in phase differences between the carrier and each sideband. This leads to a phase difference between the two beat RF signals and hence a power degradation of the ultimate RF output. In addition, high-order dispersion introduces variations in time delay between adjacent wavelength channels (*i.e.*, $\Delta t$ in **Eq. (5)**), which also leads to phase errors. For the BPD, since the intensity noise is largely cancelled out by using a balanced detection scheme, the error mainly results from the phase noise arising from the shot noise [77]. In addition, the uneven transmission response across the wavelength channels can affect the processing accuracy by introducing differences in amplitude response between the taps.

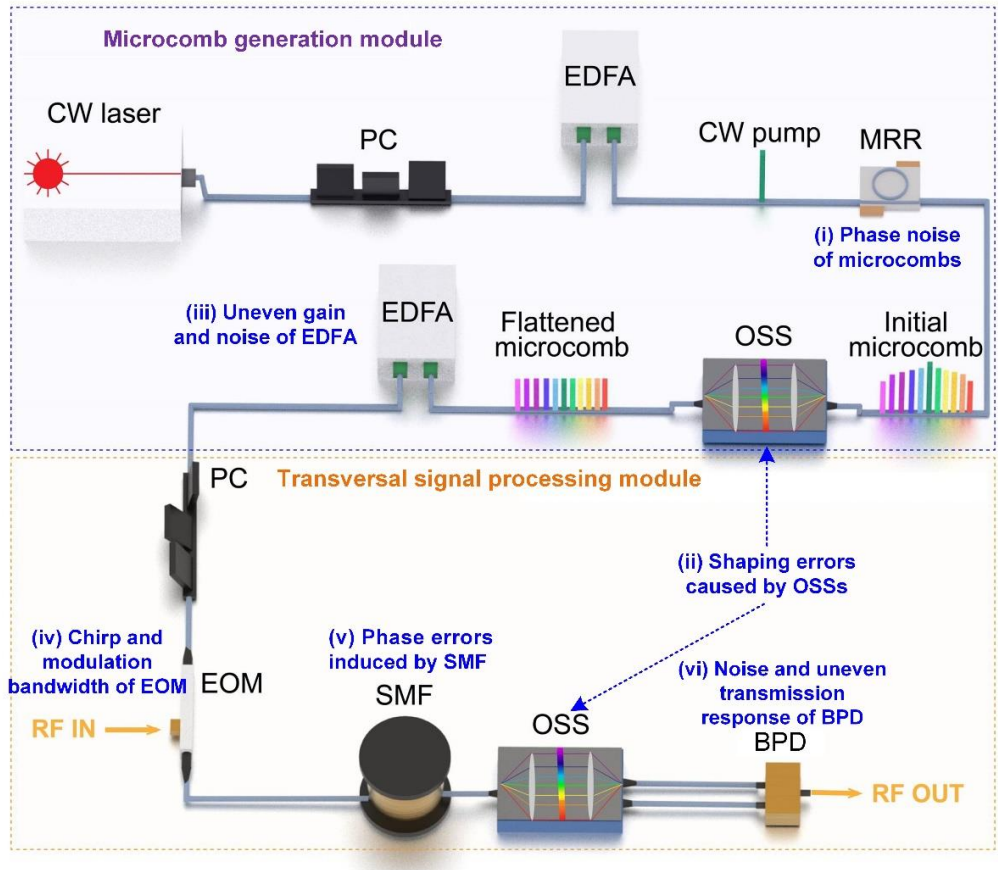

**Fig. 12.** Error sources in practical systems that affect the processing accuracy of microcomb-based photonic RF transversal signal processors, mainly including (i) phase noise of microcombs, (ii) shaping errors caused by optical spectral shapers (OSSs), (iii) uneven gain and noise of erbium-doped fibre amplifier (EDFA), (iv) chirp and modulation bandwidth of electro-optic modulator (EOM), (v) phase errors induced by single mode fiber (SMF), and (vi) noise and uneven transmission response of balanced photodetector (BPD).

The processing errors induced by the imperfect response of the EDFA, OSSs, EOM, SMF, and BPD can be reduced by introducing feedback control to calibrate the impulse response of the transversal signal processor, which we have demonstrated in our previous work [17, 31, 35, 36]. The calibration for the amplitude errors of the non-ideal impulse response can be achieved by measuring the replica of the input RF signal at each wavelength channel to obtain accurate RF-to-RF weights, followed by comparing them with the desired channel weights to generate an error signal that programs the attenuation of the OSS. To minimize the amplitude errors, several iterations along the entire feedback loop can be performed. On the other hand, the calibration of the phase errors can be realized by employing the phase modulating capabilities of the OSS to compensate the deviation between the measured and desired phase response. Recently, novel self-calibrating photonic integrated circuits have been demonstrated [78, 79], where the impulse response calibration was achieved by incorporating an optical reference path to establish a Kramers-Kronig relationship and then calculate the amplitude and phase errors based on a Fourier transform. This offers new possibilities for realizing accurate feedback control in microcomb-based photonic RF transversal signal processors. Finally, we note that, in addition to errors in the bulk instruments including EDFA, OSSs, EOM, dispersive module, and BPD in **Fig. 12**, errors also are generated for integrated components in on-chip microcomb-based photonic RF transversal signal processors that have been demonstrated very recently [80-103].

**TABLE I. COMPRISON OF THEORETICAL AND MEASURED RMSES FOR MICROCOMB-BASED PHOTONIC RF TRANSVERSAL SIGNAL PROCESSORS**

| Signal processing function [a] | Theoretical RMSE [b] | Measured RMSE without feedback control | Measured RMSE after calibrating the impulse response [c] |
|---|---|---|---|
| Differentiation | 0.23% | 3.53% | 1.15% |
| Integration | 0.71% | 4.84% | 1.77% |
| Hilbert transform | 3.96% | 7.86% | 5.09% |

[a] We chose $1^{st}$-order differentiation, integration, and Hilbert transform that corresponds to $N = 1$ in **Eqs. (6) − (8)** for comparison.
[b] We used the measured waveform of input RF signal (which has a Gaussian-like pulse waveform with a FWHM of ~0.2 ns) to calculate the theoretical RMSE.
[c] We introduced two-stage feedback control as those in Refs. [17, 31, 35, 36] to calibrate the non-ideal impulse response of the processors.

## V. THEORETICAL VERSUS EXPERIMENTAL RMSES

As mentioned in Sections III and IV, the accuracy of microcomb-based photonic RF transversal signal processors is affected by both theoretical limitations and imperfect component response, and the relative contribution of the two to the overall processing performance is of significant interest. In this section, we investigate this question by performing both theoretical calculations and experiments to compare the theoretical and measured RMSEs of microcomb-based photonic RF transversal signal processors.

**Table I** summarizes the calculated and measured RMSEs for three different signal processing functions including differentiation, integration, and Hilbert transform. In our experiments, we employed an arbitrary waveform generator (AWG) to generate the input RF signal that had a Gaussian-like pulse waveform with a FWHM of ~0.2 ns. Soliton crystal microcombs generated by a MRR made from high-index doped silica glass (Hydex) were used as the multi-wavelength source, and 80 comb lines in the C-band with a comb spacing of ~49 GHz were chosen as the discrete taps. To minimize the error between the ideal and experimentally generated Gaussian pulses, we used the input RF signal waveform measured by a high-bandwidth real-time oscilloscope in our calculations. For the measured RMSEs, we show the values with and without using feedback control to calibrate the non-ideal impulse response. Similar to our previous demonstrations [17, 31, 35, 36], a two-stage feedback control including two feedback loops that covered all the components of the processors was introduced to calibrate the non-ideal impulse response of the entire system, and three iterations were performed for each processing function.

As expected, in **Table 1** the measured RMSEs are higher than the corresponding theoretical RMSEs for all three signal processing functions. This further confirms that on the basis of theoretical limitations additional processing errors are induced by the imperfect response of the practical system. On the other hand, the measured RMSEs without calibrating the impulse response are higher than those measured after calibration that approach the theoretical RMSEs. This highlights the improvement in accuracy enabled by introducing feedback control to reduce the errors induced by experimental factors.

## VI. CONCLUSON

In summary, we thoroughly analyze the processing accuracy of microcomb-based photonic RF transversal signal processors resulting from both theoretical limitations and experimental factors. We investigate the theoretical limitations determined by the tap number, signal bandwidth, and pulse waveform, and also discuss errors induced by the imperfect response of different experimental components. Finally, the relative contributions of the theoretical limitations and experimental factors to the overall processing inaccuracy are analyzed. Our results provide a useful guide for improving the processing accuracy of microcomb-based photonic RF transversal signal processors that are versatile for realizing a wide range of high-speed processing functions for many applications.